\documentstyle[emulateapj]{article}
\begin{document}

\title{Velocity dispersions and cluster properties in the SARS
(Southern Abell Redshift Survey) clusters.
\footnote{Based on observations collected at The Las Campanas Observatory of the Carnegie Institution of Washington.} Paper II}

\author{Hernan Muriel \altaffilmark{2,3}, Hernan Quintana \altaffilmark{4}, Leopoldo Infante \altaffilmark{4}, Diego G. Lambas \altaffilmark{2,3} \& Michael J. Way \altaffilmark{4,5}}

\altaffiltext{2}{Grupo de Investigaciones en Astronom\'{\i}a Te\'orica y Experimental (IATE). Observatorio Astron\'omico, Laprida 854, 5000 C\'ordoba, Argentina.}
\altaffiltext{3}{CONICET, Buenos Aires, Argentina.} 
\altaffiltext{4}{PUC}
\altaffiltext{5}{NASA Ames Research Center, Space Sciences Division}

%------------------------------------------------------------------

\begin{abstract}
We report an analysis of the dynamical structure of clusters of galaxies
from a survey of photometric and spectroscopic observations 
in the fields of southern Abell Clusters.
We analyze the galaxy velocity field in extended regions up
to $ 7 h^{-1}$ Mpc from cluster centers and we
estimate mean velocity dispersions and their radial dependence.
Only one from a total number of 41 Abell clusters does not correspond to
a dynamically bound system. However, four of these bound objects are 
double clusters.   We estimate that
20 \% (7 clusters) of the 35 remaining are subject to serious projection 
effects. 
Normalizing the clustercentric distances by means of the overdensity radius $r_{200}$,
and the velocity dispersion profiles (VDPs)
 by the corresponding mean cluster velocity dispersion, we computed the average
 VDP. Our results indicate a flat behavior of the mean VDP
at large distances from the cluster center. Nevertheless, we found that for the inner part
of the clusters ($r/r_{200}\leq 1$) the VDP is up to a 10\% smaller than at larger radii.

\end{abstract}


\section{INTRODUCTION}

Analysis of large scale structure formation may greatly
benefit from studies of the dynamics of clusters of galaxies.
Measurements of galaxy velocity dispersions in clusters
provide reliable estimates of cluster masses and a direct normalization
of the primordial mass power spectrum (Eke, Cole \& Frenk, 1996).
Moreover, the velocity field in the extended halos of clusters
may set additional important constraints to the formation of structure
as well on the mean density parameter of the universe.

There have been several recent studies on the dynamics of clusters of
galaxies,  see for instance
Girardi et al (1993), Zabludoff et al (1993),
Collins et al (1995), Mazure et al (1996), Fadda et al. (1996)
Alonso et al. (1999).
The resulting distribution function of velocity
dispersions from the ESO Nearby Abell Cluster Survey (ENACS) given
by Mazure et al (1996) is in agreement with the distribution of
cluster X-ray temperatures, suggesting $\beta=\sigma \mu m_h/( k T_X) \simeq 1$. 
The velocity dispersion profiles (hereafter VDP) may provide
a useful tool for the study of the dynamics of clusters of
galaxies.
The analysis by Fadda et al. 1996 is consistent with a tendency of
flat VDP in rich Abell clusters.
  Jing and B\"orner 1996 investigated 
the VDPs of clusters for several cosmological models. They found that on average, VDPs
decrease with the cluster radius in every model up to $1h^{-1} Mpc$ from the cluster
center. Also, these  authors found that the slope of the profiles are different in different models, being 
steeper in lower-$\Omega$ models than in higher-$\Omega$ models.

In the hierarchical scenario of structure formation, galaxy systems
grow by aggregation of smaller structures formed earlier. Therefore,
we expect a significant degree of substructure in clusters of galaxies if
the remnants of the accretion of groups in the recent past has not been
erased by the dynamical relaxation of the clusters.
The substructure in rich clusters has been extensively analyzed in recent years
(Dressler \& Shectman (1988), West \& Bothun (1990), Zabludoff et al. (1993), 
Girardi et al. (1997), Solanes et al. (1999)).
The results are consistent with substructure in most of the cases studied,
irrespective of the samples and method of analyses used.
West \& Bothun, (1990) made an analysis of substructure in clusters
of galaxies and their surroundings. The authors developed a technique
that is sensitive to correlations between galaxy positions and local
kinematics finding little evidence for substructure in the inner regions,
and significant departures from a relaxed substructure-free systems
in the external regions.  More recently, Biviano et al. 2002 realized
a detailed analysis of the consequences of substructure on luminosity and morphological
segregation. These authors find that the number of galaxies in substructures decrease markedly toward the cluster center and report differences in the properties of 
galaxies depending whether they belong to substructures or not. These differences are also 
present in the the dynamical properties of galaxies.

Escalera et al. (1994) provide an extensive discussion of the presence of
substructure in clusters of galaxies by using galaxy positions and redshifts.
In their studies a multi-scale analysis is adopted that considers the kinematics
as well as the wavelet transform providing estimators of the degree of 
substructure. Other works (see for instance Fadda et al. 1996)
consider velocity gradients and anisotropy of galaxy orbits.
Extensions of the different
methods of analysis can provide new useful quantitative estimates of 
substructure, essential for a better understanding of the dynamics of 
clusters of galaxies.

The dynamics of clusters of galaxies 
may also be studied from information in the X-ray band.
X-ray emission detected in a large fraction of clusters of galaxies
provides an invaluable observational material.
Several properties of the clusters and the intra-cluster medium 
may be addressed with this information. For example, the global mass
distribution, the dynamical state and the evolution with redshift
, the composition of the intra-cluster medium, etc.
White (1999) presents an elegant methodology to recover
the spatial properties of the intra-cluster gas from X-ray observations.
From the deconvolution of ASCA satellite X-ray data, he finds a
large fraction (90\%) of clusters consistent with isothermality. 
These results are
in conflict with the Markevitch et al. (1998) analysis from a sample of 30
clusters where most show
steeply-declining intra-cluster temperature profiles.
In their analysis of ASCA resolved spectroscopic data these authors
obtained projected temperature profiles and in many cases two-dimensional
temperature maps concluding that the gas temperature varies within a factor
1.3-2 or greater within the clusters.

The conflicting evidence for isothermality of the intra-cluster medium
show that
 the information on the VDP for clusters may add important information
to the subject. On the other hand, the tendency for subclustering
to occur at large distances from cluster centers encourages us to explore
the outer regions of clusters of galaxies. 
In this paper, we analyze the radial velocity distribution in regions
extending up to $7 h^{-1}$Mpc  in projection from the Abell cluster center 
(we adopt $H_0$=100$hkms^{-1}$ and $q_0$ =0.5).
We provide a detailed analysis of each individual cluster providing
the degree of substructure and an estimate of the VDP at large distances
from the cluster center. In section 2 we describe the method of analysis
of substructure. Section 3 deals with the identification of
the clusters and the projection effects which significantly affect 
the measurements of velocity dispersions. Section 5 provides the
estimates of mean velocity dispersion and the correlation with
richness counts as well as the velocity dispersion profile of several clusters.
%The data sample is described in paper I (Way et al. 2002).

%----------------------------------------------------------

\section{Data}

The SARS survey  (Southern Abell clusters Redshift Survey, Way et al. 2002) comprises  
Abell 1958 and Abell, Corwin \& Ollowin 1989 (hereafter ACO)
clusters R $\geq$  1, principally in the region 0 $\leq \delta \geq$-65 
and 5h$\geq \alpha \leq$    21h (avoiding the
LMC and SMC), with b $\leq$   -40.
Galaxies were selected from the APM catalog (Maddox el
al. 1990). Galaxies brighter than $m_R =19$    and with
surface brightness    within 1.5 x 1.5 $deg^2$ centered
on the cluster were pre-selected. 
Target galaxies were selected at random and the 
final completeness is roughly constant up to an apparent magnitude 
$\sim 18$ and it is of the order of 75\%.

The observations were carried out with the 2.5m DuPont
telescope at the Las Campanas Observatory, Chile. The multi-fiber
spectrograph (Shectman
1989) was used. 
Fibers are connected to a Boller \& Chivens
spectrograph attached to a 2DFrutti detector (2DF). 
The unknown spectra were calibrated using
software packages within IRAF,  following essentially
the method described by Way et al. (1997).

From the wavelength calibrated spectra was obtained the respective radial
velocities cz of the unknown spectra by using two
different, independent methods: i) The Fourier Cross Correlation
Technique, where two fourier transformed spectra, the unknown
object and a known template, are multiplied together to obtain
the fourier transform of their correlation function (with RVSAO,
Tonry et al. 1979). ii) "By eye" identification of
absorption lines and computation of cz with the task {\it rvidlines}.
The final sample consists in more than 4000 galaxies with redshift estimates in 41 
clusters. Mean cluster's redshift run from 0.06 to 0.16 with a mean 
around 0.088.

%---------------------------------------------------------

\section{Algorithm for substructure detection}

We have applied  two different techniques in order to detect
 substructure in clusters. These techniques are complementary in the sense
that they are mainly designed to remove large structures along
the line of sight, and smaller systems in three dimensions.
Many clusters present double structures in the redshift distribution (e.g  late 
stage of a cluster-cluster merger). Ashman et al. (1994)  discuss a statistical 
technique for detecting and quantifying bimodality known as mixture
modeling or  the KMM algorithm.
The scheme is based on the application of algorithms that fit 
a certain number of substructures in redshift space  to the data   
and it is determined the best fitting model. 
This technique is the base of 
commonly adopted procedures used to analyze astronomical data sets
and it assesses the statistical significance of bimodality providing 
objective ways of dividing the data into sub-populations.
As discussed by Ashman, Bird \& Zepf (1994) the KMM technique has broad
applicability in the analysis of astronomical data. 
We have applied this technique to the redshift distribution regardless the 
angular position of the galaxies in the field of the cluster.
Based on a preliminary inspection of the data we propose the number of 
structures with  their corresponding mean radial velocities and velocity 
dispersion that approximately represent the redshift distribution around 
the cluster. This procedure is restricted only to those structures with 
overlapping redshift distributions. Then we  apply  the KMM algorithm and
consider a multiple peak structure when the  confidence level to have
proposed model  is bigger than 90\% (for details see Ashman et al. 1994).
We have considered different possibilities: i) when the proposed model for
multiple-peak-structure is rejected we consider  that the redshift distribution
corresponds to a single cluster; ii) if the 
confidence level of the proposed model is bigger than the 90\% and
at least 70\% of the galaxies belong to the same  
structure we assume a single cluster and discard the outlying groups
 that will deserve a detailed study in a future work.; iii) when most 
of the galaxies belong to
two separate structures of similar sizes we assume the presence of two
clusters and we perform the corresponding analysis. It should be noted that
the above technique works properly when the sub-structures 
are representative of an important number of galaxies. In this work only 
structures with at least 10 galaxies are considered. 
  
In the hierarchical model for large scale structure  formation,  clusters of 
galaxies are the result of a continuous process of accretion of small
structures like groups of galaxies. Therefore, a considerable number of
galaxies are expected to be found around clusters that are not bound to
the main system  and therefore will bias the velocity dispersion 
estimate if they are included in the analysis. This problem is particularly 
serious if large distances from the cluster center are considered as is
the case in the present work.  Some of these groups of galaxies can be
located at a similar redshift to the cluster, therefore, they are very
difficult to detect in the redshift space .

In our data set, and for each cluster, we analyze the real nature of visually 
identified group candidates with a technique that considers both the projected
position and redshift of the galaxies using the compactness of the
projected distribution  and the departure from the mean dynamical properties
of the cluster. Three different parameters are used:

i) a $\delta$ parameter similar to that defined by Dressler and Shectman 1988:

$$ \delta^2=(11/\sigma^2)[(v_{local}-v)^2+(\sigma_{local}-\sigma)^2] $$

where  $\sigma$ and $\sigma_{local}$ are the velocity dispersion of the 
cluster and the proposed group structure 
respectively and $v$ and $v_{local}$ are the corresponding mean velocities.

ii) A parameter $C$, that provides a measure of the compactness of groups and
is computed as: $C= <dnn_{local}>/<dnn_{group}>$ where $<dnn_{group}>$
is the average projected distance  of the nearest neighbors  to each of the ng
members of the proposed group  and $<dnn_{local}>$ is computed 
in the same way but for the nearest ng galaxies in the neighborhood of the 
proposed group. 

iii) An isolation parameter $I=dnng/<dnn_{group}>$ 
where $dnng$ is the distance to the nearest  neighbor galaxy to the group.

We compute the variable $G=\delta + C + I$ and we calculate 
the mean  $\langle$G$\rangle$ and the dispersion $\sigma_G$ for each cluster.
A given  group candidate  
is to be removed if the value of G for the group is at least 2
standard deviations, $2 \sigma_G$, away from the cluster mean value $\langle$G$\rangle$.
The adopted threshold is the result of  Monte Carlo simulations which show
that this threshold is suitable to remove structures. For the five
most regular clusters in our sample we reassigned the
polar angle of every galaxy with respect to the cluster center. 
This procedure removes  group structures and
leaves unchanged the radial galaxy density profile of the cluster. For the 
mock data, 
we compute $G$ identifying mock  groups of galaxies
finding that none of these chance groups have  G-$\langle$G$\rangle \geq  
2\sigma_G$ (with $\langle G \rangle$ and $\sigma_G$  computed from  the 
original data).

In spite of the fact that our sample of galaxies in clusters is not 
magnitude limited the above procedure should give no biased results 
provided the galaxies are randomly selected from a complete sample.

%----------------------------------------------------------

\section{Analysis}

As a result of the two techniques described above we have removed 19 structures in
14 clusters from our  total sample of 41 Abell clusters. We find that
 4 Abell clusters appear as two different systems  in
redshift space, and one Abell cluster is completely spurious. 
In the following sections 
we discuss different properties of the resulting 44 clusters. Table 1 shows the Abell number; the total number of galaxies with measured redshift in the line of sight of
the cluster ($N_{tot}$); the number of galaxies assigned to the cluster ($N_{clu}$)
and the cluster mean  radial velocities. 
 
\subsection{Cluster identification}

In our analysis we have only considered clusters selected by Abell 1958 and ACO. 
Several authors (van Haarlem et al. 1997 and references therein)
have discussed the consequences of the projection effects when clusters
are selected in 2-D catalog. Redshift surveys provide precise information on the
reality of the clusters selected. As a result of our analysis we found only
one spurious cluster (Abell 3159) while the rest appear as real concentrations
in redshift space.  Nevertheless, 11 clusters present more than one
concentration in the redshift space, thus in projection they appear as richer
clusters. Of the total of 40 Abell clusters 28 appear as a single concentration
in the total redshift range, while the rest have been systematically 
enhanced  by projection effects.  We consider that a cluster is significantly affected by 
projections when the  number of galaxies in groups or other
cluster-like structures along  the line of sight 
 is comparable to the number of confirmed cluster members. 
Besides the projection effects, and after the removal
of structures previously described, several clusters present
significant evidence for substructure on different levels. This 
substructure can affect the analysis of the
dynamics of clusters. In particular, the estimate of the
velocity dispersion may be significantly affected by substructure. 

We will use the term ``relaxed" cluster to describe a system that
is free from substructure with a single nearly gaussian redshift distribution 
after the subtraction of structures using the procedure discussed in Section 3.
This classification will be used to define sub-samples of clusters. Our original
cluster sample is not volume  complete,
and the above definition is used to select
subsets of clusters  in order to cross-correlate general properties such as richness counts, mean 
velocity dispersions, etc. Figures 1a and 1b show the mean redshift distribution of 
the total sample clusters and those classified  as "relaxed" respectively where 
it can be appreciated the similarity of the redshift distributions.

%------ FIG 1
\placefigure{fig-1}
\def\figureps[#1,#2]#3.{\bgroup\vbox{\epsfxsize=#2
    \hbox to \hsize{\hfil\epsfbox{#1}\hfil}}\vskip12pt
    \small\noindent Figure#3. \def\par{\endgraf\egroup\vskip12pt}}
\figureps[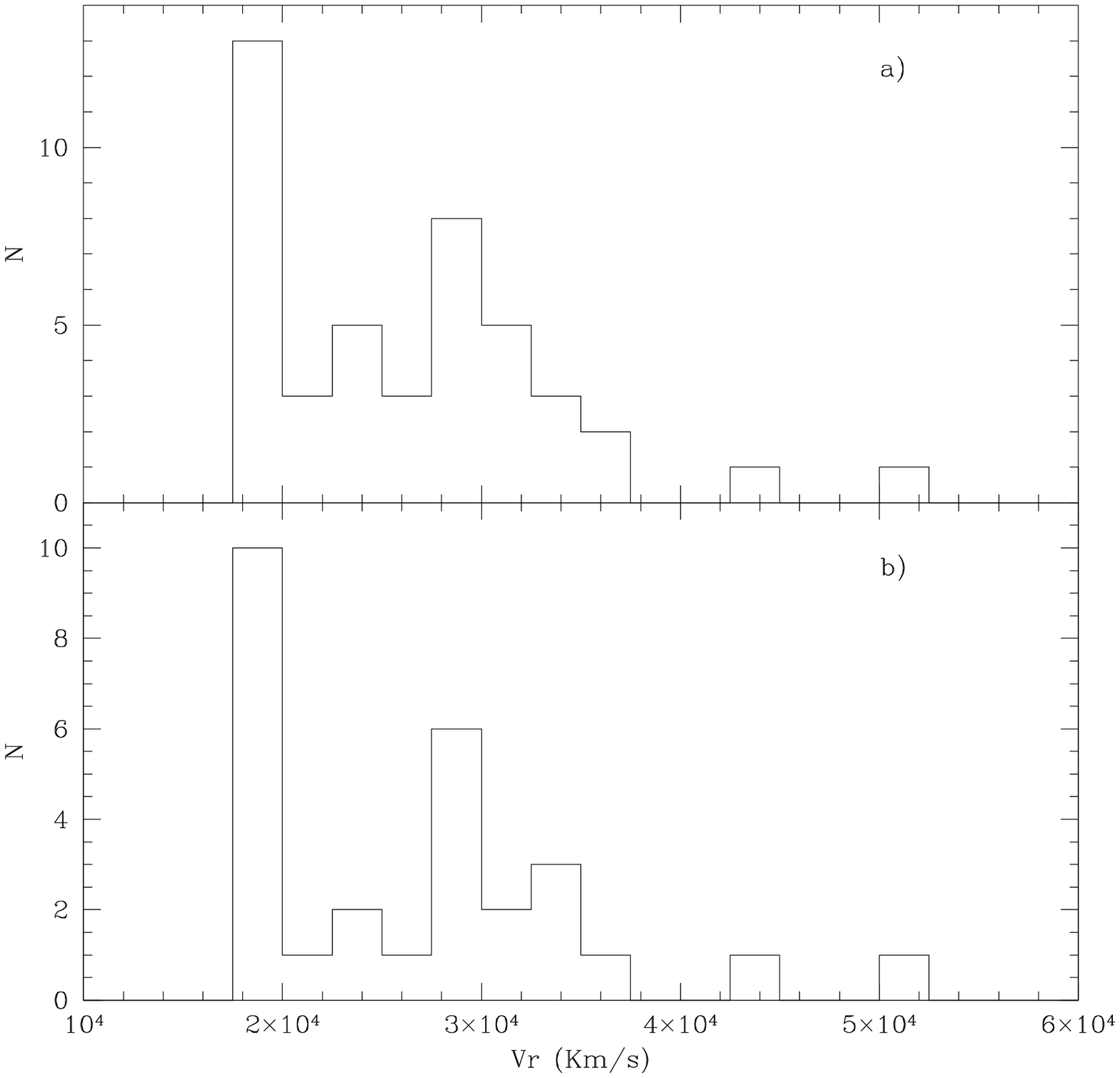,1.00\hsize] 1. 
Histogram of mean radial velocities of the clusters:
a) the total sample; b) "relaxed" clusters.
%-------

%-------------------------------------------

\subsection{Substructure properties}

We have analyzed the  properties of the different structures removed from 
clusters. As a result of  the algorithm of group detection we find an average 
velocity dispersion $\sigma_g = 295 \pm 180 km/sec$ and a mean extension $D=0.44 
\pm 0.28  h^{-1} Mpc$. These structures which comprise 12\% of the total number 
of galaxies in the clusters  have, on average, 
a difference of mean velocity respect to the parent cluster  
$\Delta V = 921  \pm 393 km/sec$.
Our values of $\sigma$ and $D$ are typical of group of galaxies. Nevertheless,
  the average extension of our groups are larger than those derived by  Girardi et al. 1997 
($\sim 0.2 h^{-1} Mpc$).
The KMM technique  for substructure detection tends to identify  systems at larger distances
from the cluster center ($\Delta V = 1515 \pm 304 km/sec$). Nevertheless, the mean velocity 
dispersion of these structures ($288 \pm 128 km/sec$) is similar to the $\sigma_{g}$ 
derived by the group detection algorithm.

%--------------------------------------------------

\subsection{Individual objects}

Several clusters in our analysis deserve individual attention due to different
peculiarities of their properties.

The redshift distribution along the line of sight of A2819, A2871, A3107 and
A3223 show two similar structures not physically connected. In Table 1
each of these clusters are named with the original Abell number plus
an ``a" and ``b" respectively.

As an example of the application of the method we comment on the cluster A2734
which presents a double peaked structure in redshift space.  The
smallest peak has approximately half the number of members of the main
structure.  It was removed since the probability to have two different
structures is 99 \%. 

A380 presents some evidence of a double structure with a mean
velocity difference of 1407 km/s. The probability to have two
different structures is larger than 90\%.  Nevertheless, the low
number of galaxies (25) involved  in our
analysis introduce some doubts into our conclusions. The values quoted
in Table 1 correspond to a single cluster. Assuming two different
structures we find the following values:
$ \langle V \rangle =31440$ km/s, $\sigma$=408 for the nearest structure  
(14 galaxies) and 
$<V>=32847$ km/s, $\sigma$=314 for the second (11 members).    

Besides the clusters that appear as double in the redshift space, A380,
A487, A2915, A3142, A3153, A3844, A3864 present strong projection
effects due to the presence of several structures such as groups
of galaxies along the line of sight.
  
A3111: This cluster shows some evidence of large scale substructures
in the plane of the sky. Our algorithm does not work properly for this type
of substructure,  therefore, the cluster was taken as a single structure and  
the value quoted in Table 1 (943 km/s)  could be biased  high.
Nevertheless, the discrepancy with the velocity dispersion derived by
Fadda et al. 1996 (159 km/s) can not be explained. If we arbitrarily restrict 
ourselves to the central region of the cluster (up to 2.5 Mpc/h in diameter)
where  no evidence
of substructure is present we derive a $\sigma=734 km/s$. This value must
be taken as a lower limit for the mean velocity dispersion of A3111.
The value derived by Fadda et al. 1996 is probably biased by the low
number of confirmed members in their sample (12 galaxies)
while our analysis is based  on more than 50 cluster members.

A3151 presents a group of galaxies in the very center of the cluster
with a mean velocity differing by more than $900$ km/s with respect
to the main cluster.  This is  nearly  the same difference  between
our estimate of the cluster mean velocity and the value derived by
Fadda et al. 1996. Their estimate is the result of 14 galaxies, and by 
chance they selected galaxies from the projected group instead of
the main cluster.

A3223 appear as two separate structures in the plane of the sky and
hence was treated as two different clusters. These two clusters also
show important differences in their dynamical properties.
The second concentration  was identified by the APM selection
criteria and is named as APMCC 479.

After the removal  of groups and besides those indicating double structures
A1750, A3111 A3135, A3764  A3915 still present some evidence for
substructure in redshift space or in the plane of the sky.

A3159: The   redshift distribution  in the line of sight of this cluster  
shows the presence of several groups, nevertheless, none of these  groups
can by classified as a cluster. We suggest this system is a spurious
cluster identification.

A2778  and A3153 are two clusters  poorly defined  both in the plane of
the sky and in the redshift space where the presence of gaps suggest
the possibility of substructure. More redshifts are needed in order
clearly understand these clusters.  The values of $\sigma$ quoted
in Table 1 for these two clusters must be taken with caution,
especially in the case of A3153 where the redshift distribution can be also
consistent with several groups instead of a single cluster.

\section{Velocity dispersion estimates}

After the redefinition of structures as defined in Section 2 
we have computed the mean velocity dispersion for each cluster.   
Based on the  ROSTAT routine (see Beers et al. 1990) we have used robust
mean and scale estimators. We have applied relativistic corrections and
have taken into account velocity errors. Considering the typical
number of redshift confirmed cluster members (usually $>$ 20)
we have considered the {\it biweight }
estimate for both the mean cluster radial velocity and the velocity
dispersion. Errors are based on the statistical {\it jacknife}. The derived values are
shown in Table 1.  Figure 2a shows the values of the mean velocity dispersion for clusters  in the range 200 - 1100 km/s with a mean $\approx$  600 km/s
indicating that in our study we have included low mass systems 
(probably groups) as well as massive clusters of galaxies.
Figure 2b shows the same distribution but only for those clusters 
classified as ``relaxed". As it 
can be appreciated no differences are present 
between both sets of data indicating that contamination by projection
effects are seen 
at some degree in all clusters, irrespective of redshift and $\sigma$.

%------------ FIG 2
\placefigure{fig-2}
\def\figureps[#1,#2]#3.{\bgroup\vbox{\epsfxsize=#2
    \hbox to \hsize{\hfil\epsfbox{#1}\hfil}}\vskip12pt
    \small\noindent Figure#3. \def\par{\endgraf\egroup\vskip12pt}}
\figureps[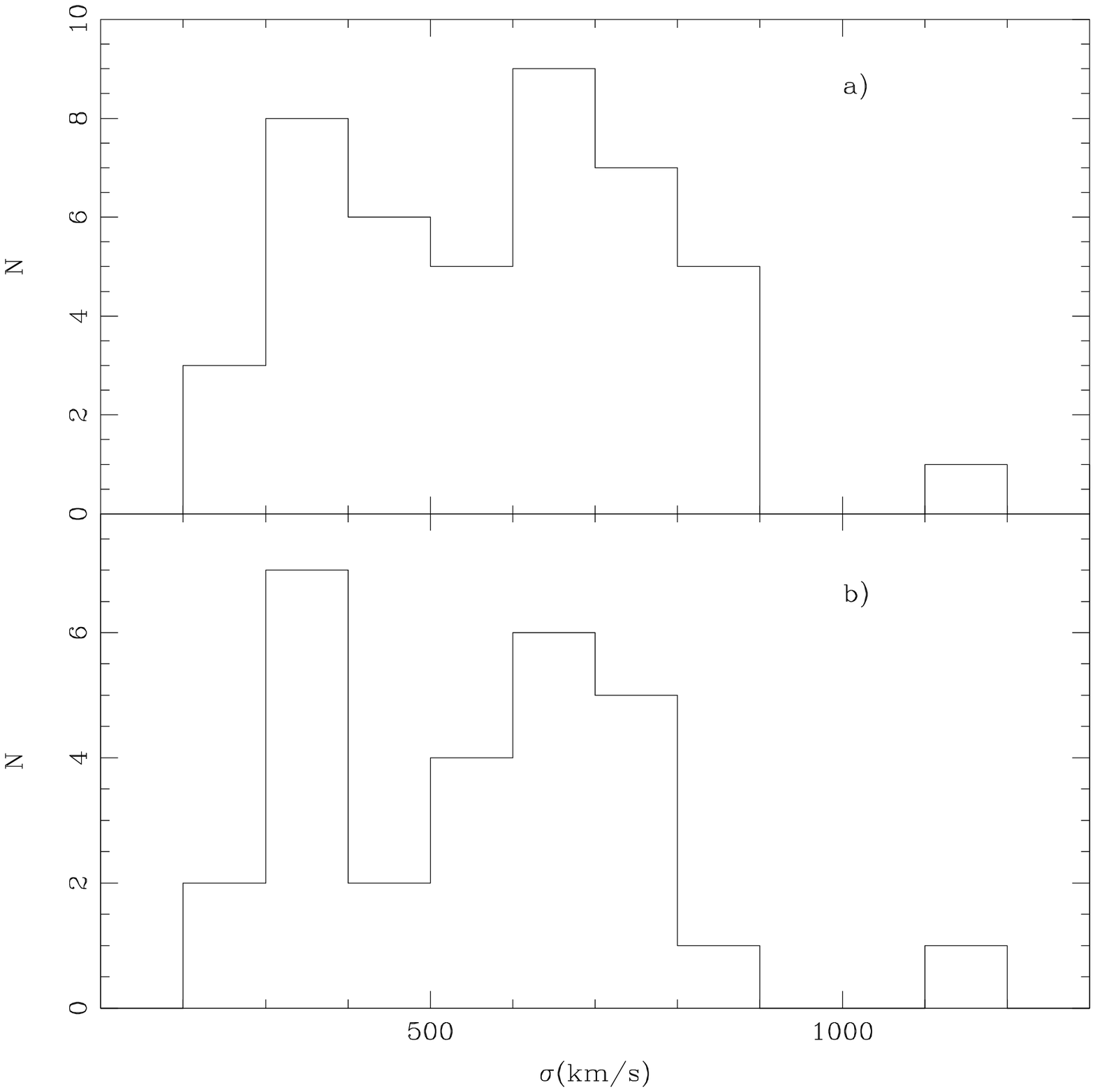,1.00\hsize] 2. 
Histogram of mean velocity dispersion of the clusters: a) the
total sample; b) "relaxed" clusters.
%--------------

\subsection{Comparison with other estimates}

20 of the clusters in our sample are also in the ENACS survey. Figure 3a
shows the comparison between our estimates and those obtained by
Fadda et al. 1996. We found a mean difference $<\sigma_{Fadda}-\sigma_{SARS}>=-89\pm132$ that indicate that
our values of $\sigma$ are on average  slightly higher than those in 
Fadda et al. 1996.
If we restrict our sample to those clusters  
with at least 30 confirmed members (the same restriction is applied for 
Fadda et al. 1996)  we  find  $<\sigma_{Fadda}-\sigma_{SARS}>=-40\pm108$ 
which suggest a smaller shift and spread (see figure  3b).
In both cases we have made the comparison assuming the same cluster 
radius as  Fadda et al. 1996 (typically smaller than our maximum cluster radii).

%----------- FIG 3
\placefigure{fig-3}
\def\figureps[#1,#2]#3.{\bgroup\vbox{\epsfxsize=#2
    \hbox to \hsize{\hfil\epsfbox{#1}\hfil}}\vskip12pt
    \small\noindent Figure#3. \def\par{\endgraf\egroup\vskip12pt}}
\figureps[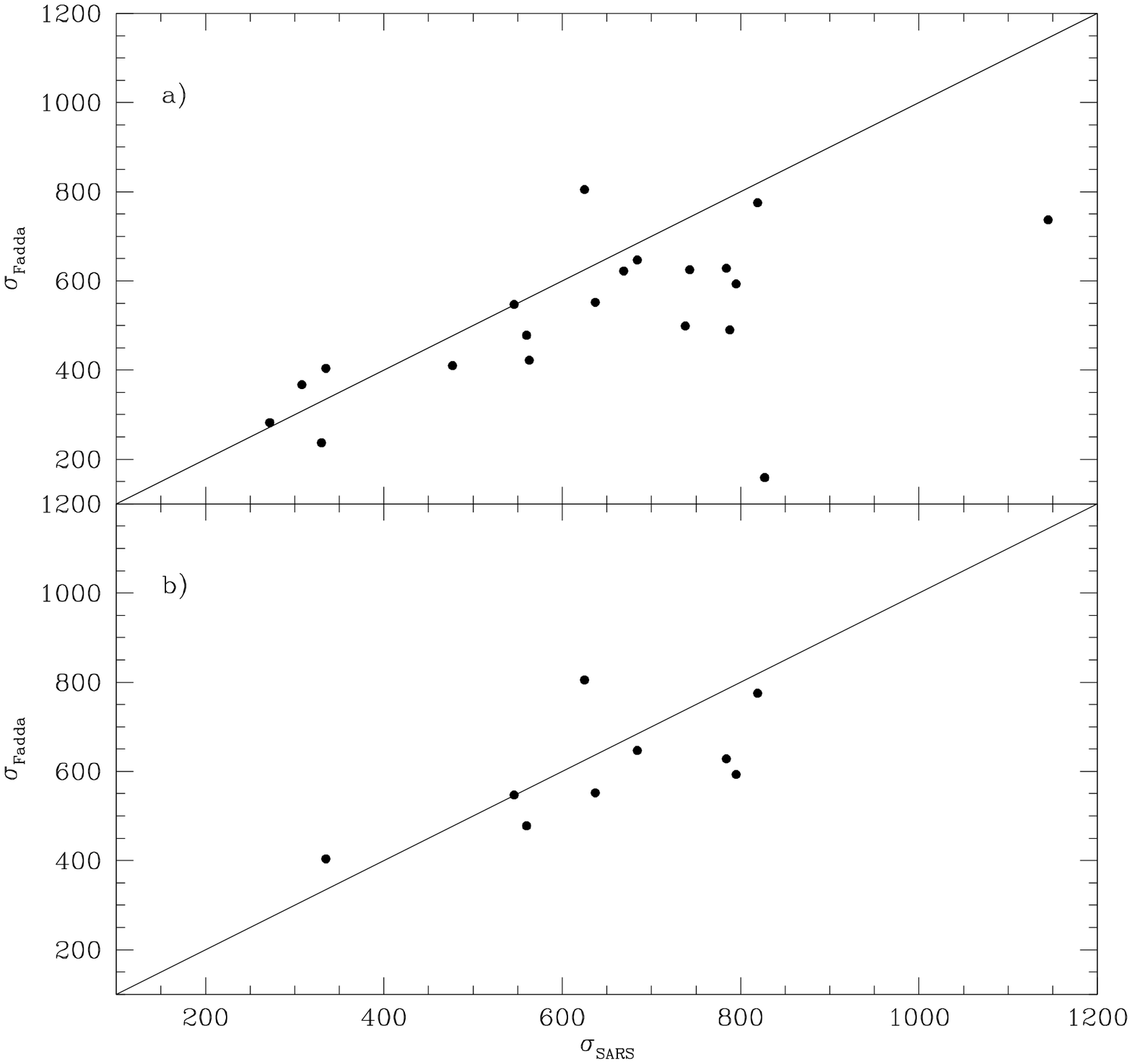,1.00\hsize] 3. 
{a) Comparison with Fadda et al. 1996 results. The solid
line corresponds to equal $\sigma$. b) Same as a) for  clusters with
at least 30 confirmed members in both sample. The solid line corresponds
to equal $\sigma$.
%--------------

\subsection{Cluster mean velocity dispersion vs. richness counts}

Taking into account the methods previously described, we have computed
the mean velocity dispersions for our sample of clusters.
We have performed a comparison between $\sigma$ and the richness number counts
$\cal N$ as defined in the ACO catalog.  Since only a small fraction
of our cluster sample are known X-ray emitters, we have not
attempted to analyze correlations between our dynamical estimates and
the X-ray information.

Figure 4a show the correlation between $\sigma$ and richness counts
$\cal N$ taken from the ACO catalogue where no clear correlation can be 
appreciated. A similar result was found by Mazure et al. (1996). These
authors suggest that the very broad relation between $\cal N$  and $\sigma$ 
must be largely intrinsic.
Nevertheless, when we restrict to the ``relaxed" clusters
and ``isothermal" distributions (gaussian velocity distribution and VDP
flat or slowly decaying, see the next section for details) 
and exclude those clusters more strongly affected by  projection effects, 
the data suggest some    
correlation between richness counts and $\sigma$ in the sense 
that richest clusters tend to 
have higher $\sigma$.  This correlation can be see in  Figure 4b, where
a linear fit has been applied deriving the following relation:
$\sigma = (6.2\pm 2.8) {\cal N} + (158 \pm 202)$. 

%----------- FIG 4
\placefigure{fig-4}
\def\figureps[#1,#2]#3.{\bgroup\vbox{\epsfxsize=#2
    \hbox to \hsize{\hfil\epsfbox{#1}\hfil}}\vskip12pt
    \small\noindent Figure#3. \def\par{\endgraf\egroup\vskip12pt}}
\figureps[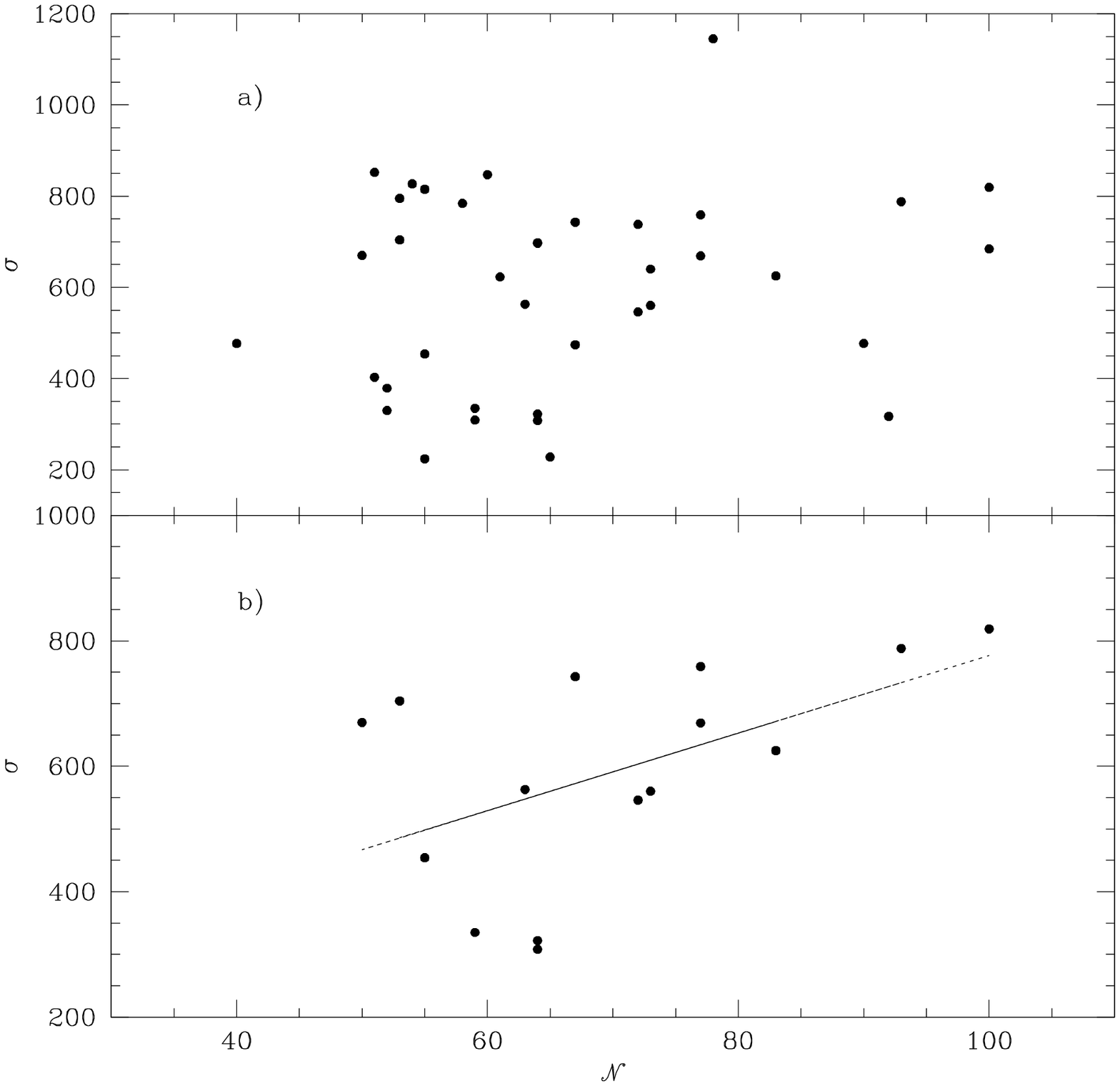,1.00\hsize] 4. 
a) Correlation between the mean velocity dispersion estimate
and the richness counts estimated by ACO. b) Same as a) for "relaxed" clusters.
%------------

\subsection{Velocity dispersion profiles}

The large projected area around clusters in the SARS survey allows
us to analyze of the dynamics of galaxies in the extended
halos of clusters. A useful statistical measure of the dynamics
is the velocity dispersion profile, the velocity 
dispersion at a given radius evaluated by using all the galaxies
within that radius.  The VDP was computed for the 29 clusters in 
our sample with more than 20 confirmed members.  We have used a step  size 
of 0.5 Mpc $h^{-1}$ while most VDP are computed up to 4 Mpc $h^{-1}$ 
and in some cases beyond  this radius.  Many clusters
show an irregular trend in the inner part (cluster
radius $\leq$ 1 Mpc $h^{-1}$), this effect could be partially related to
the low number of galaxies in the inner part of the cluster and may
also depend on the choice of the cluster center which in our case
correspond to the values provided by Abell.
Nevertheless, the most interesting aspect of the VDPs is the behavior at large
distances from the cluster center.  For these 29 clusters we find 19  (14 are  
``relaxed" clusters) with a  flat VDP,  5 that present a slowly decaying 
profile and  5 with a rising profile. It should be noted that only 1 of
the VDP rising clusters was  classified as  ``relaxed" cluster. 
These results are shown in Figure 5 for the 29 objects with reliable estimates
of VDP.

\newpage

%----------- FIG 5
\placefigure{fig-5}
\def\figureps[#1,#2]#3.{\bgroup\vbox{\epsfxsize=#2
    \hbox to \hsize{\hfil\epsfbox{#1}\hfil}}\vskip12pt
    \small\noindent Figure#3. \def\par{\endgraf\egroup\vskip12pt}}
\figureps[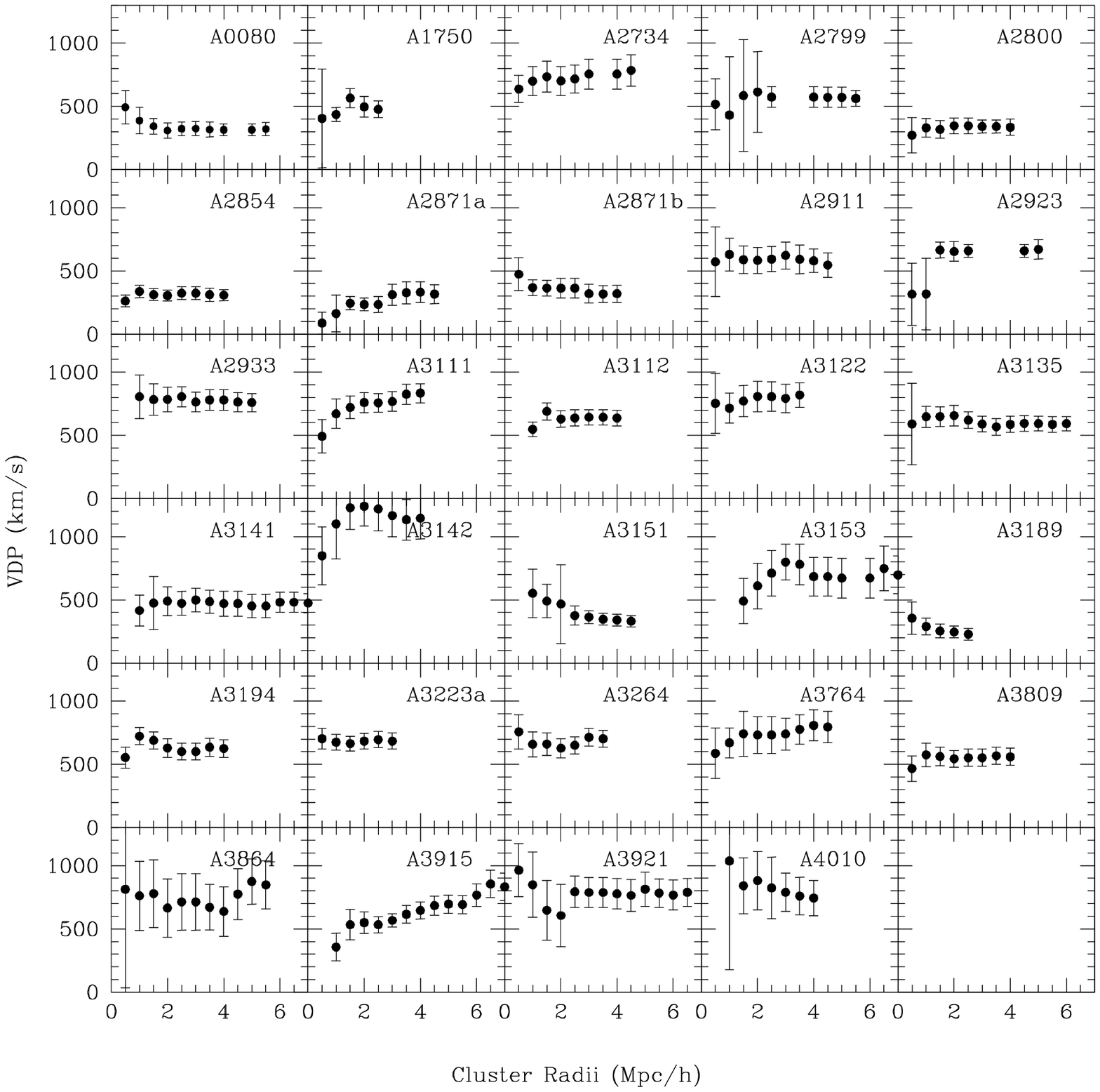,2.0\hsize] 5. 
Velocity dispersion profiles for 29 Abell clusters.
%-------------

\newpage

In order to allow for a
physical comparison between clusters with different mean velocity
dispersions, we have normalized cluster's radii using $r_{200}$
(the radius where the mean interior cluster overdensity is 200).  
Assuming a singular isothermal  profile  Carlberg et al.
1997 derive the following correlation
between $r_{200}$ and the cluster mean velocity dispersion: 
$ r_{200} = \frac{\sqrt{3} \sigma}{10 H_0(z)}$. We have followed a similar 
analysis than that  
proposed by den Hartog and Katgert (1996) and Jing and B\"orner (1996) 
consisting in the computation of the ratios of $\sigma$ at different distances 
from the cluster center. Both den Hartog and Katgert (1996) and Jing and 
B\"orner (1996) use the radius in $Mpc$ (up to 3 $Mpc h^{-1}$ and 1 
$Mpc h^{-1}$ respectively).  
We propose the use of a normalized radius and  
four bins for the computation of the 
velocity dispersion estimates $\sigma_1$,
$\sigma_2$,  $\sigma_3$ and $\sigma_4$:  $r/r_{200}<1$, 
$r/r_{200}<2$, $r/r_{200}<3$,  $r/r_{200}\leq 7$ respectively.
The shape of the VDP at different radii can be quantified by the ratios  
$\sigma_i/\sigma_j$ (=1 for a flat profile). 
In figure 6 we show the distribution of the following ratios: $\sigma_1/\sigma_2$,
$\sigma_2/\sigma_3$,  $\sigma_1/\sigma_3$, $\sigma_1/\sigma_4$ 
$\sigma_2/\sigma_4$ and the derived mean values are 0.93, 1.00,
0.95, 0.96 and 1.00 respectively.
It can be appreciated that those ratios involving
$\sigma_1$ suggest that in the inner bin (up to $r/r_{200}=1$) the 
velocity dispersion is approximately 10\% lower than at larger distances.
This fact can also be appreciated in figure 7 where the total
sample of clusters has been averaged  being each VDP profile normalized with 
the corresponding mean cluster velocity dispersion. Figure 7 also clearly shows 
that the average  VDP for $r/r_{200}>1.5$ is nearly flat  well beyond 
the cluster virial radius.

%--------- FIG 6
\placefigure{fig-6}
\def\figureps[#1,#2]#3.{\bgroup\vbox{\epsfxsize=#2
    \hbox to \hsize{\hfil\epsfbox{#1}\hfil}}\vskip12pt
    \small\noindent Figure#3. \def\par{\endgraf\egroup\vskip12pt}}
\figureps[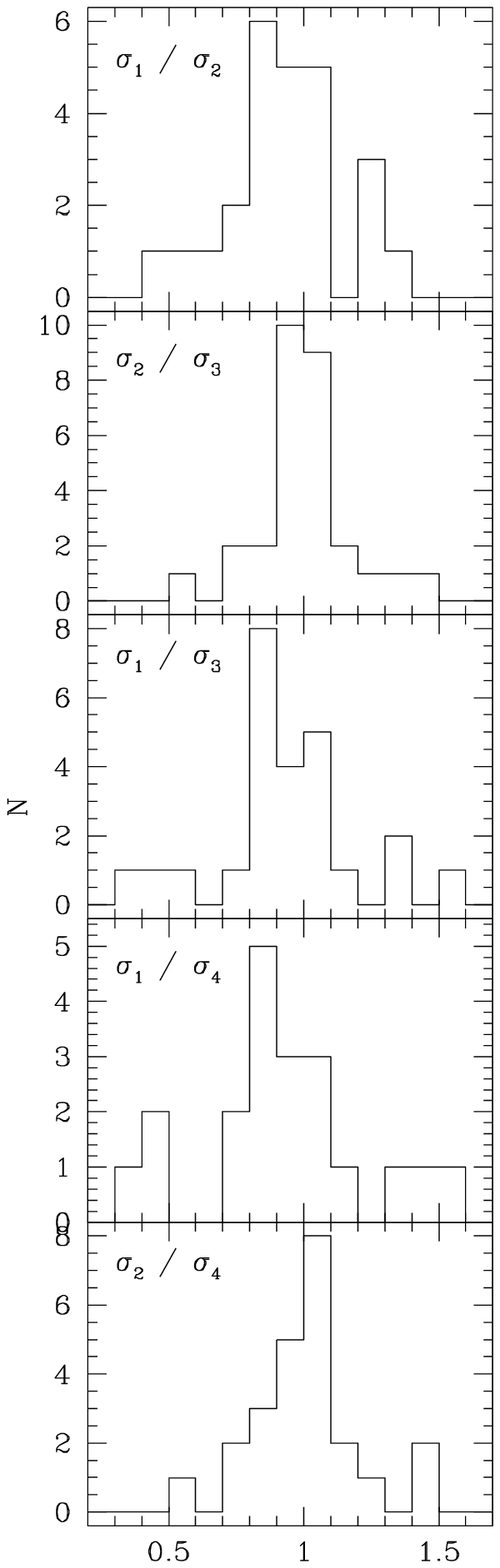,1.40\hsize] 6. 
Velocity dispersion ratios between $\sigma$s computed for different
normalized bins:  $\sigma_1$, $\sigma_2$,  $\sigma_3$ and $\sigma_4$ correspond
to $r/r_{200}<1$, $r/r_{200}<2$, $r/r_{200}<3$,  $r/r_{200}\leq 7$ respectively.
%-----------

%--------------------------------------------------

\section{Discussion and Conclusions}

A significant degree of substructure in clusters of galaxies 
is expected in the hierarchical scenario of structure formation.
This is because of the large time-scale for the remnants of the
accretion of groups onto clusters in the recent past to be 
erased by dynamical relaxation.  On the other hand, the
identification of clusters in two dimensions may be strongly biased by 
spurious systems due to projection effects as shown in numerical simulations 
(van Haarlem, Frenk \& White 1997).
These two issues heavily complicate a detailed analysis of the 
dynamical properties of clusters of galaxies.

The evidence for substructure in rich clusters have been extensively explored
in different studies of the galaxy distribution in cluster fields where the 
most accepted view is the relevant presence of substructure.
The X-ray observations also contribute to our knowledge of the
spatial properties of the intra-cluster gas.
However, recent analyses  provides conflicting results
regarding the isothermality of the gas or the existence of 
steeply-declining temperature profiles.
This issue, and the fact that the degree of substructure
increases at large distances from cluster centers motivated the present study 
of the outer regions of clusters of galaxies.  Our work is mainly centered
in the analysis of the radial velocity distribution of galaxies in
extended regions of Abell clusters, focusing on the existence of gradients
in the velocity dispersion profiles.

We have carried out an analysis of the velocity field of galaxies in
extended regions up to $7 h^{-1}$ Mpc from the cluster centers.
We have applied several methods to remove 
contamination by projection effects and analyzed the presence of sub-clustering.
We have obtained suitable estimates of the mean velocity dispersions
and its radial dependence using the ROSTAT routines. Our analysis can
be compared to  Fadda et al. 1996 for a fraction 
of common objects. It is clear from our analysis that the larger 
differences arise in those clusters
with more contamination and a smaller number of measured redshifts.
We also find that the correlation between mean velocity dispersion
$\sigma$ and richness number counts $\cal N$ is strongly affected by
projection effects. There is some evidence of  correlation between $\sigma$ and
$\cal N$ for a sub-sample restricted to systems with no significant 
degree of contamination. 

From our original sample of 41 Abell clusters we found that 40 are real
clusters although 4 of these appear as double systems.
These results are similar to those found by Mazure et al. 1996. These 
authors found that almost all ACO clusters with richness class 1 or greater 
correspond to real systems in the redshift space and about 10\% of the ACO 
clusters appears to be the result of a superposition of two similar poorer 
systems.  Beside  the double systems we also found that  7 of the 
clusters in our sample are
subject to serious projection effects. 
The  fraction of clusters
with a high degree of contamination in our sample compares well with the results of such
effects in the mock catalogs from the numerical simulations of
van Haarlem, Frenk \& White 1997) where 1/3 of Abell-type clusters
are expected to arise from projection of groups along the line of sight.

From the resulting sample of 44 clusters, four are poorly defined and more data are
needed in order to better establish their properties. In spite of 
projection effects, 28 of the 44 clusters are well defined in
redshift space and have a velocity distribution consistent with a 
gaussian function.

Our results show that the average VDP is flat at large distances from
the cluster center. This behavior is found for 19 clusters (65\% of the 29 
with VDP estimates) indicating that  an isothermal hypothesis can be assumed
even at radii well beyond the virial radius. Nevertheless, we found that on average, the
normalized velocity dispersion 
is about 10\% smaller in the inner region
of the clusters ($r/r_{200} \leq 1$). 
A possible interpretation for the decay of VDPs in central regions can be 
related to the morphological segregation in clusters. Fadda et al. 1996 
found kinematical segregation in the sence that early type galaxies show smaller 
values of $\sigma$ than late types and Mazure et al. 1996 found that the  
brightest cluster galaxies (typically of early type morphology) move  
slower than other galaxies and Ram\'{\i}rez et al. 1998  found that 
the differences between the velocity distributions of elliptical
and spiral galaxies are associated with the shape of their orbit families. 
Since early type objects are dominat 
within $r_{200}$ the results shown in figure 7 are to be expected. 

%--------- FIG 7
\placefigure{fig-7}
\def\figureps[#1,#2]#3.{\bgroup\vbox{\epsfxsize=#2
    \hbox to \hsize{\hfil\epsfbox{#1}\hfil}}\vskip12pt
    \small\noindent Figure#3. \def\par{\endgraf\egroup\vskip12pt}}
\figureps[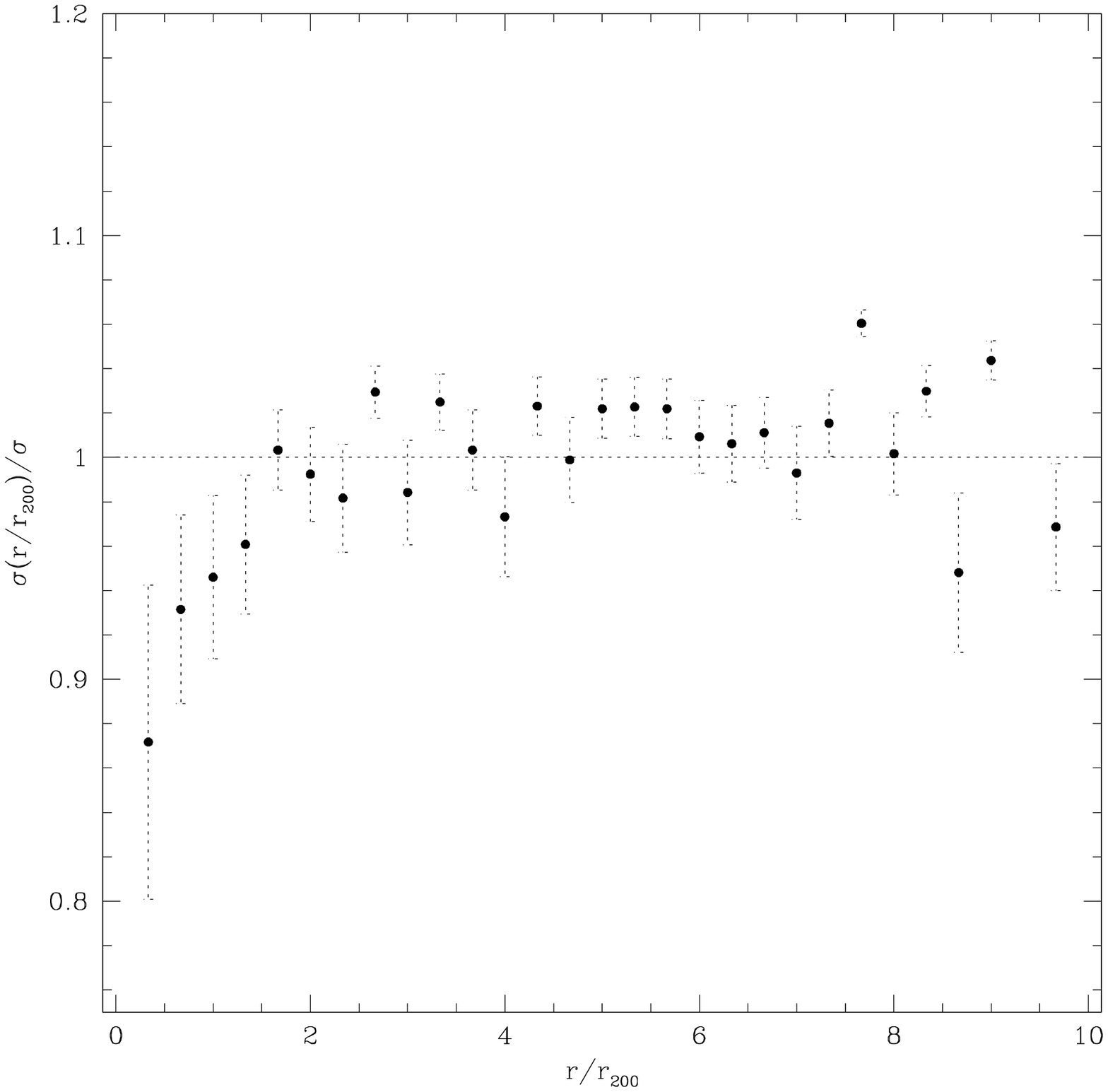,1.00\hsize] 7. 
Mean VDP for the total sample of clusters with VDP estimate. Each
cluster has been normalized using the corresponding mean $\sigma$.
%-----------

The shape of VDP profiles are of fundamental importance for their 
implications on cluster properties and cosmology since detailed theoretical 
predictions from different cosmological scenarios could be used to
set restrictions to current models of structure formation. Jing and B\"orner 1996 
analysis of  
VDPs of clusters for several cosmological models show an average decline of VDPs
with the distance to cluster centers.  
Nevertheless, this analysis was restricted to 
the very inner  region of the clusters ($\leq 1h^{-1} Mpc$. Therefore, new numerical simulations 
must be analyzed to test our findings of flat VDPs at very large distances from the cluster
center. 

Under the assumption that clusters are in global dynamical equilibrium
(even beyond the virialization radius) our results can be compared to 
temperature radial distributions derived from the X-ray emission of the 
intra-cluster medium.  Flat VDP at large clustercentric distances
may shed light on the recent  controversy  on the nature, either
flat or  declining, of intracluster temperature radial profiles (see White 2000).

\noindent Acknowledgments

We thank the Referee for useful comments and suggestions which
greatly improved the original version of this paper.
This work was partially supported by the Consejo de Investigaciones
Cient\'{\i}ficas y T\'ecnicas de la Rep\'ublica Argentina, CONICET;
SeCyT, UNC and Agencia C\'ordoba Ciencia, Argentina.
L. Infante and H. Quintana acknowledge FONDECYT and  FONDAP, Chile for 
partial support. H. Quintana acknowledges the support of a Catedra 
Presidencial en Ciencia, Chile.

%----------------------------------------------------------

%----------------------------------------------------------
\end{document}